\documentclass{article}

\usepackage{graphicx}
\usepackage[final,nonatbib]{neurips_2023_ml4ps}
\usepackage[numbers]{natbib}

\usepackage[utf8]{inputenc} 
\usepackage[T1]{fontenc}    
\usepackage{hyperref}       
\usepackage{url}            
\usepackage{booktabs}       
\usepackage{amsfonts}       
\usepackage{nicefrac}       
\usepackage{microtype}      
\usepackage{xcolor}         
\usepackage{amsmath}
\usepackage{caption}
\usepackage{subcaption}
\usepackage{soul}

\newcommand{\Mpch}{\,h^{-1}\text{Mpc}}

\title{Probabilistic reconstruction of Dark Matter fields from biased tracers using diffusion models}

\author{%
  Core Francisco Park \\
  Harvard University \\
  17 Oxford St., Cambridge, MA 02138, USA \\
  \texttt{corefranciscopark@g.harvard.edu} \\
  \And
  Victoria Ono \\
  Harvard University \\
  17 Oxford St., Cambridge, MA 02138, USA \\
  \texttt{victoriaono@college.harvard.edu} \\
  \AND
  Nayantara Mudur \\
  Harvard University \\
  17 Oxford St., Cambridge, MA 02138, USA \\
  \texttt{nmudur@g.harvard.edu} \\
  \And
  Yueying Ni \\
  Harvard-Smithsonian Center for Astrophysics \\
  60 Garden Street, Cambridge, MA 02138, USA \\
  \texttt{yueying.ni@cfa.harvard.edu} \\
  \AND
  Carolina Cuesta-Lazaro \\
  The NSF AI Institute for Artificial Intelligence and Fundamental Interactions \\
  Massachusetts Institute of Technology, Cambridge, MA 02139, USA \\
  \texttt{cuestalz@mit.edu} \\
}

\begin{document}

\maketitle

\begin{abstract}
Galaxies are biased tracers of the underlying cosmic web, which is dominated by dark matter components that cannot be directly observed. 
The relationship between dark matter density fields and galaxy distributions can be sensitive to assumptions in cosmology and astrophysical processes embedded in the galaxy formation models, that remain uncertain in many aspects.   
Based on state-of-the-art galaxy formation simulation suites with varied cosmological parameters and sub-grid astrophysics, we develop a diffusion generative model to predict the unbiased posterior distribution of the underlying dark matter fields from the given stellar mass fields, while being able to marginalize over the uncertainties in cosmology and galaxy formation. 
\end{abstract}

\section{Introduction}

The nature of dark matter (DM) remains one of the most enigmatic questions in cosmology. Direct observation of dark matter remains elusive. N-body simulations have in the past \citep{Jasche_2019} been used to unveil the dark matter density field from tracers such as galaxies. Over the last decade, significant progress has been made in cosmological hydrodynamic simulations of galaxy formation \citep[e.g.][]{pillepich2018simulating,Dave2019MNRAS.486.2827D,Bird-Astrid}. These simulations study how galaxies evolve over cosmic time and allow more accurate exploration of the relationship between observed galaxy distributions and underlying dark matter fields. 
For example, previous work of Hong et al. \cite{hong2021revealing} developed a convolutional neural network model based on the IllustrisTNG simulation, using the stellar density and velocities to regress the otherwise unobservable dark matter  density maps.
However, machine learning models trained solely on a single simulation can not accommodate the substantial uncertainties inherent in the astrophysical processes assumed by galaxy formation models, as well as the specific cosmological parameters employed in that simulation. Moreover, it is important to develop probabilitic models, such as \cite{mudur2022denoising}, that can describe our uncertainties in the way that galaxies connect to the underlying dark matter distribution to answer questions such as what is the likelihood of a certain dark matter halo mass associated with a given galaxy.

In this work, we develop a diffusion generative model trained based on CAMELS simulation suites \cite{villaescusa2021camels,Ni2023arXiv230402096N} that contain more than 1000 state-of-the-art galaxy formation simulations with varied cosmological parameters and sub-grid astrophysics, to reconstruct the underlying DM fields from stellar fields.
The primary goal of the model is to capture the relationship between the stellar fields and DM fields by marginalizing over the uncertainties in the modeling of galaxy formation, and predict the unbiased posterior distribution of the DM fields conditioned on the given stellar field, $p(x_\mathrm{DM}|x_\mathrm{stars})$.

\bibliographystyle{unsrt}

\begin{figure}[h]
  \centering
\includegraphics[width=0.8\textwidth]{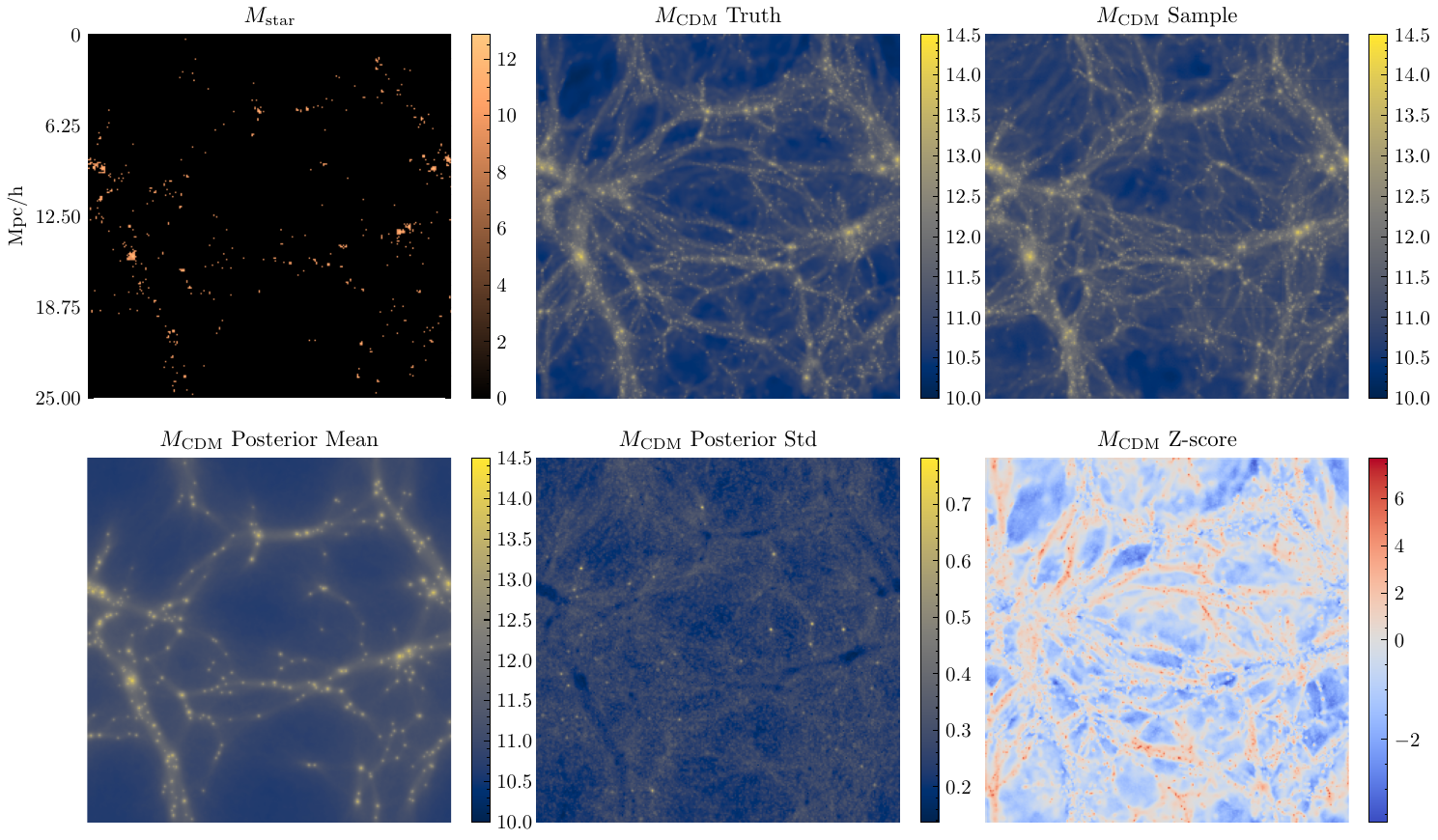}
  \caption{\textbf{Top row:} Input stellar  field, corresponding true DM, and a sample DM from the diffusion model. \textbf{Bottom row:} Posterior mean, posterior standard deviation, and pixelwise $z$-score from 100 posterior samples.}
  \label{fig:fields}
\end{figure}

\section{Methodology}
\paragraph{Model}
We use the variational diffusion model developed by Kingma et al. \cite{kingma2021variational} with a denoising architecture similar to the U-Net \cite{ronneberger2015u} to model the posterior of DM density fields given stellar fields. The conditional diffusion model generates a target DM density field in $T=250$ refinement steps. It begins with a random noise field $x_\mathrm{DM}^T \sim \mathcal{N}(0,1)$, and iteratively denoises it according to a learned conditional probability $p_\theta(x_\mathrm{DM}^{t-1}|x_\mathrm{DM}^{t}, x_\mathrm{stars})$ to ultimately generate a sample  $x_\mathrm{DM}^0 \sim p(x_\mathrm{DM}|x_\mathrm{stars})$. The denoising model therefore takes as input $\{x_\mathrm{DM}^{t},x_\mathrm{stars}, t\}$ and estimates the noise that was added to the image at that timestep. In the forward diffusion process, we progressively add noise to an image by sampling from 
$q(x_\mathrm{DM}^{t} \vert x_\mathrm{DM}^{0}) = \mathcal{N}(\alpha_t x_\mathrm{DM}^{0}, \sigma_t^2\mathbf{I})$,
where $\alpha_t$ and $\sigma^2_t$ are functions of $\gamma_t$, which is assumed to be a linear function of time and whose free parameters are learned during training. Noise is added to the sample in a variance-preserving way, i.e. $\alpha_t^2 = \text{sigmoid}(-\gamma(t))$. The loss function we optimize is the variational lower bound of the marginal likelihood. 

\paragraph{Dataset}
For training, we use the Latin Hypercube set of the IllustrisTNG suite from the 2D CAMELS Multifield Dataset \cite{villaescusa2022camels} at $z=0$. This contains 1000 simulations sampling a wide variation range in cosmological and astrophysical parameters, reflecting the uncertainties of cosmology and the complex astrophysical processes taking place in our current understanding of galaxy formation. This training set includes 15 samples for each of the 1000 simulations, where each simulation is uniquely varied by the parameter values of $\Omega_\text{m}, \sigma_8$ (cosmological), $A_{\text{SN1}}, A_{\text{AGN1}}, A_{\text{SN2}}$, and $A_{\text{AGN2}}$ (astrophysical). The range of parameters is: $0.1 \leq \Omega_m \leq 0.5, 0.6 \leq \sigma_8 \leq 1.0, 0.25 \leq
(A_{\rm SN1}, A_{\rm AGN1}) \leq 4.00$, and $0.5 \leq (A_{\rm SN2}, A_{\rm AGN2}) \leq 2.0$. 
The samples are 2D projections of $5 \Mpch$ thickness from the 3D density fields of stars and dark matter. 
We evaluate our model on the one-parameter (1P) set, which contains 11 variations of a single parameter at a time. We take the log of each field and standardize the dataset to have zero mean and unit variance. We augment the dataset by adding random translations (with periodic boundary conditions), flips and permutations of the input and output images. We keep the image size to the original $256 \times 256$ pixels.

\paragraph{Training}

We use a U-Net \cite{ronneberger2015u} like architecture with 4 blocks of double convolution followed by strided downsampling layers. We employ group normalization \cite{wu2018group} and residual connections \cite{kaiming_resnet} in each block, and use the AdamW optimizer \cite{adamW} with a learning rate of $1\times 10^{-4}$. We also initialize the learned linear noise schedule with $\gamma(t)=18.3\:t-13.3$. We train the model using the PyTorch Lightning framework \cite{Falcon_PyTorch_Lightning_2019} with a batch size of 12 for 60000 gradient steps.

\section{Results}
We first compare the DM fields sampled from the diffusion model for the CAMELS simulations at the fiducial parameters. In the top row of Fig.~\ref{fig:fields}, we show an example stellar map together with the ground truth underlying DM map, and one of the samples from the diffusion model. The second row shows the diffusion model's posterior mean and standard deviation from  100 samples, conditioned on the same stellar field. The posterior standard deviation and the pixelwise $z$-score, computed as $\frac{\text{Truth - Posterior Mean}}{\text{Posterior Std}}$, allow us to assess whether the diffusion model's uncertainty is sufficient to account for deviations from the target field. The posterior standard deviation is lowest at the positions of structures that were present in the stellar field, while the $z$-score is higher for filaments that are present in the target DM field but do not correspond to structures in the stellar field. Note that the $z$-score is not symmetric since the overdensity field is lower bounded by $-1$.

\begin{figure}[h]
  \centering
\includegraphics[width=\textwidth]{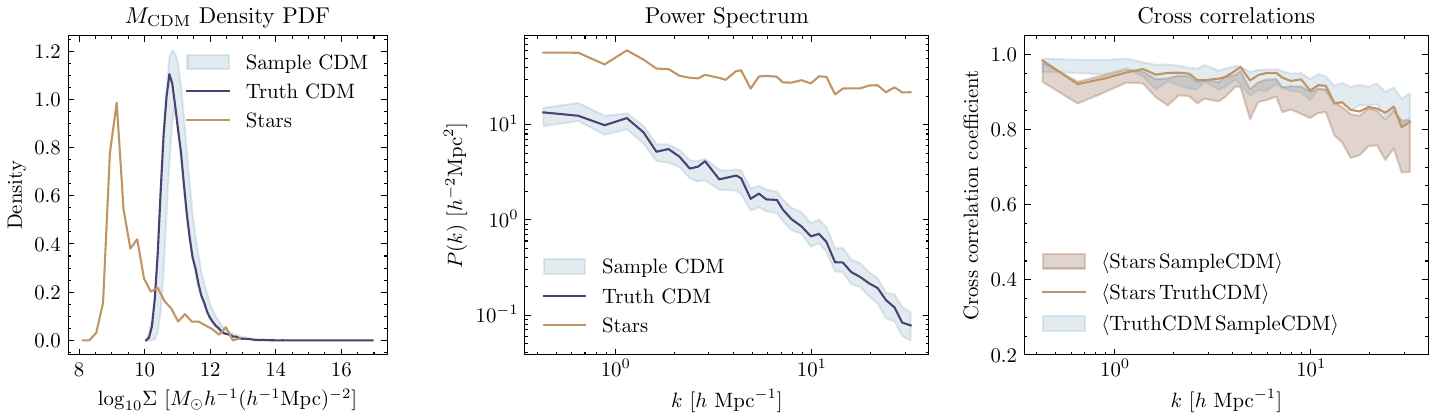}
  \caption{
  Summary statistics of the stellar and DM density field at the fiducial parameters.
  \textbf{Left panel:} density histogram of stars (copper), true DM (solid purple lines) and DM inferred by the diffusion model (light blue). Note that we show the 10-90th percentiles of $100$ samples from the posterior distribution. \textbf{Middle panel:} Power spectra for star, true DM and sampled DM fields. \textbf{Right panel:} Cross correlations between i) stellar and DM density fields, for both true and inferred DM fields (the brown line and shaded region), and ii) between true and sampled DM fields (the blue shaded region).
  }
  \label{fig:summaries}
\end{figure}

Fig.~\ref{fig:summaries} shows a quantitative comparison between the summary statistics of the true DM fields and those of the generated ones. 
We find good statistical agreement between the true DM field and the sampled DM fields, both in density histogram (the left panel) and in the clustering of the DM fields (the middle panel).

The right panel compares the statistics of the cross-correlations.

The cross-correlations between the sampled and true DM fields are always higher than $0.8$; the large-scale cross-correlation is as expected larger than that on small scales, but it is overall very well reproduced by the diffusion model samples.

\begin{figure}
    \centering
    \includegraphics[width=1.0\textwidth]{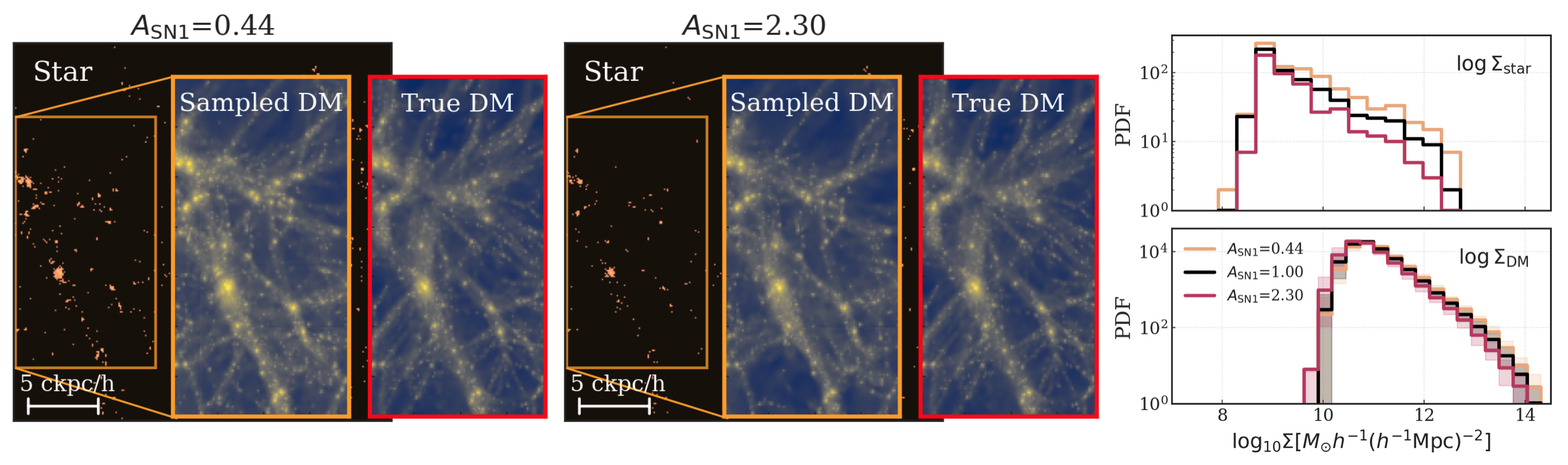}
    \caption{
    \textbf{1st and 2nd panels}: Illustrations of the DM maps generated from stellar fields from two simulations of different SN wind strength (low $A_{\text{SN1}}$ for the first panel and high $A_{\text{SN1}}$ for the second panel). The red panels show the respective true DM maps for comparison. 
    \textbf{3rd panel}: Histogram of the stellar density (top) and generated DM density (bottom) for the simulations run with SN strengths of low, fiducial and high values. The shaded regions show the 10-90th percentiles based on 100 generated samples for each of the input stellar maps. 
    }
    \label{fig:ASN1-example}
\end{figure}

To demonstrate the model's capability to marginalize over the range of cosmological and astrophysical parameters we trained on, we test the model on the 1P set of CAMELS which varies one parameter (from the fiducial value) at a time for each simulation.  Fig.~\ref{fig:ASN1-example} shows samples from the diffusion model when conditioned on stellar fields from simulations with varying strengths of supernovae (SN) feedback $A_{\rm SN1}$. 

In these simulations, higher $A_{\rm SN1}$ (stronger SN feedback) leads to significant regularization of star formation and results in a lower stellar population compared to the scenario with lower $A_{\rm SN1}$ (weaker SN feedback), as evident from the stellar image and density histogram. 
However, within both scenarios of strong and weak SN feedback, the diffusion model is able to predict the underlying DM field from the given stellar field that well resembles the ground truth DM distribution both visually and on a statistical basis.
This exhibits the capability of the diffusion model to predict the DM distribution from biased tracers of galaxies produced from different astrophysical scenarios.

\begin{figure}[h]
  \centering
  \includegraphics[width=0.95\textwidth]{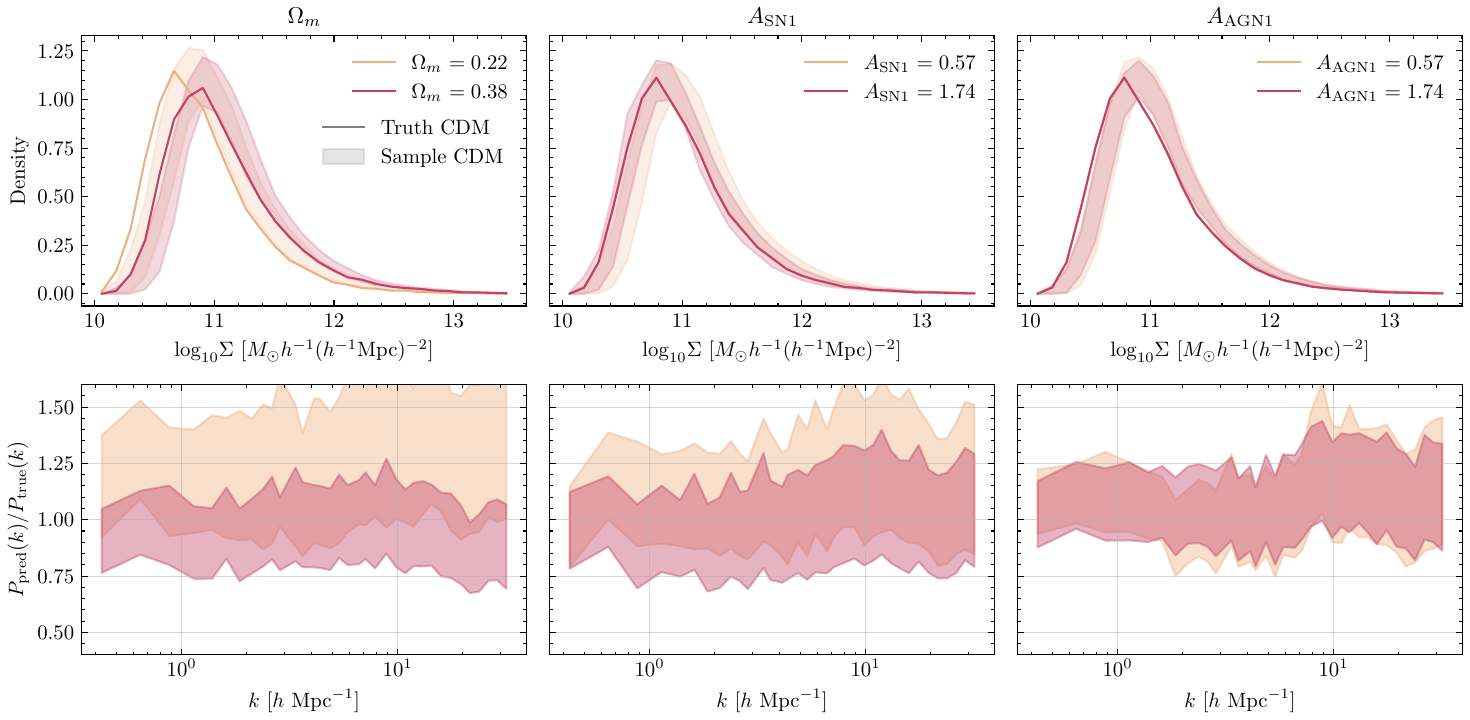}
  \caption{
  Summary statistics of the density fields when varying $\Omega_m$, $A_\mathrm{SN1}$ and $A_\mathrm{AGN1}$. The first row shows a comparison of the density field PDFs. The second row shows the power spectrum ratio between the sampled fields and the true DM distribution. The solid lines in the top panel give the true DM distributions from simulations, while the shaded regions correspond to 10-90th percentiles based on 100 generated DM samples from the input stellar fields. Note that the dark matter projected density distribution does not vary noticeably with the astrophysical parameters $A_\mathrm{SN1}$ and $A_\mathrm{AGN1}$, and the two solid lines overlap.}
  \label{fig:1P_summarize}
\end{figure}

Fig.~\ref{fig:1P_summarize} shows a quantitative comparison between the generated DM density fields and the true ones based on the 1P set simulations that separately vary the parameters $\Omega_m$, $A_{\rm SN1}$ and $A_{\rm AGN1}$.
The DM distribution is clearly sensitive to the cosmological parameter $\Omega_m$. We can see that the model is able to predict systematically lower (higher) DM distribution for the low (large) $\Omega_m$ scenario, exhibiting the potential to generalize the DM predictions across varying cosmological assumptions. However, the model struggles to generalize for low $\Omega_m$ values, predicting the DM power with amplitude larger than expected, albeit showing larger uncertainties in the samples. 
Similarly, the model also overpredicts the DM power for the low $A_{\rm SN1}$ scenarios. 
Improving the capacity of the model to generalize over the entire parameter space will be the subject of future work.

By comparing the dark matter power spectrum, we find that the predicted and observed DM fields exhibit a broad level of agreement, with a deviation of approximately $20\%$, although the uncertainty in the diffusion model is large enough for the true DM field to not be anomalous.
The consistency of the power spectrum with varied cosmological and astrophysical assumptions demonstrates that the trained model can well capture the clustering properties of the underlying DM fields based on the stellar fields, while marginalizing over the cosmological and astrophysical parameters.

\section{Conclusions and Outlook}
We have presented a diffusion generative model that can sample the posterior distribution of dark matter density fields conditioned on observable stellar maps. Previous deep learning approaches used regression-based frameworks, but we show that a generative model can be used to turn this mapping into a probabilistic one and capture the inherent uncertainty of the model due to lacking information in the stellar maps. 
The diffusion model can generate DM density fields with summary statistics and visual inspections consistent to those in the authentic simulations.
This consistency persists even when stellar maps are changed notably by variations in parameters controlling the SN and AGN feedback, illustrating the model's capacity to effectively marginalize over astrophysical uncertainties.
Our future work will extend the current framework to 3D density fields for an application to galaxy surveys and explore how additional information, such as galaxy velocities, luminosities and colors, can be folded in to better constrain the DM density field.

\section{Code Availability}
Our code is publicly available at \href{https://github.com/cfpark00/vdm4cdm}{https://github.com/cfpark00/vdm4cdm}

\bibliography{references}

\section{Supplementary Material}

\subsection{Statistical analysis of the sampled fields}
\begin{figure}
    \centering
    \includegraphics[width=\textwidth]{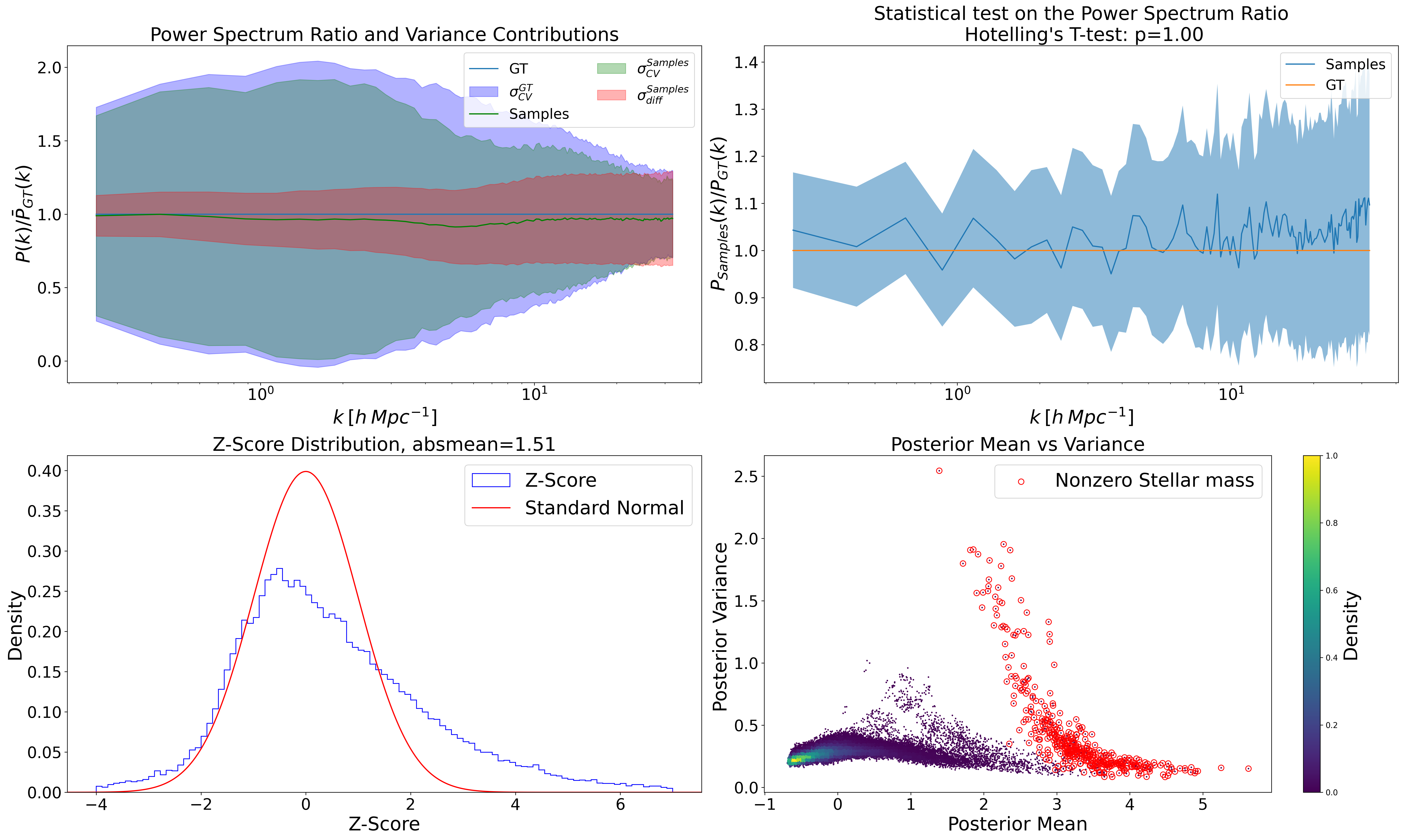}
    \caption{Statistics on 500 samples generated from an input stellar mass map from the CV (Cosmic Variance) set. This set of simulations have the same fiducial parameters but a different initial condition. \textbf{Top left:} Cosmic Variance of GT Dark Matter fields and variance from Cosmic Variance and diffusion seed of the generated samples. \textbf{Top right:} Hotelling's T-test with a diagonal covariance assumption shows that, for a single sample, the GT Dark Matter's power spectrum is consistent with the mean power spectrum of the samples. \textbf{Bottom left:} The Z-score distribution of a single GT Dark Matter field with respect to the per-pixel posterior mean and variance of generated samples. \textbf{Bottom right:} Joint distribution of the posterior mean and variance of the DM field for a single conditioning stellar mass field. Pixels corresponding to a non-zero stellar mass fields are in red circles.}
    \label{fig:supp_stats}
\end{figure}

We illustrate some more statistical properties of the sampled DM fields in Figure~\ref{fig:supp_stats}. We show the variance of the power spectrum due to cosmic variance and due to sampling the conditional posterior distribution described by the diffusion model. The latter is only bigger at the smallest scales. The results are within the expected level of variance from the ground truth DM fields. We show that the GT DM power spectrum is statistically consistent with the mean of the distribution of the sampled DM field's power spectra, and the absolute mean of the Z-score is also order of unity. Finally, we find that the pixels with non-zero stellar mass form a distinct branch (shown in red) in the joint distribution of the posterior mean and variance values. Thus, the variance is the biggest in the pixel where there is non-zero but small underlying stellar mass.

\subsection{Illustrations of the diffusion process}

\begin{figure}
    \centering
    \includegraphics[width=\textwidth]{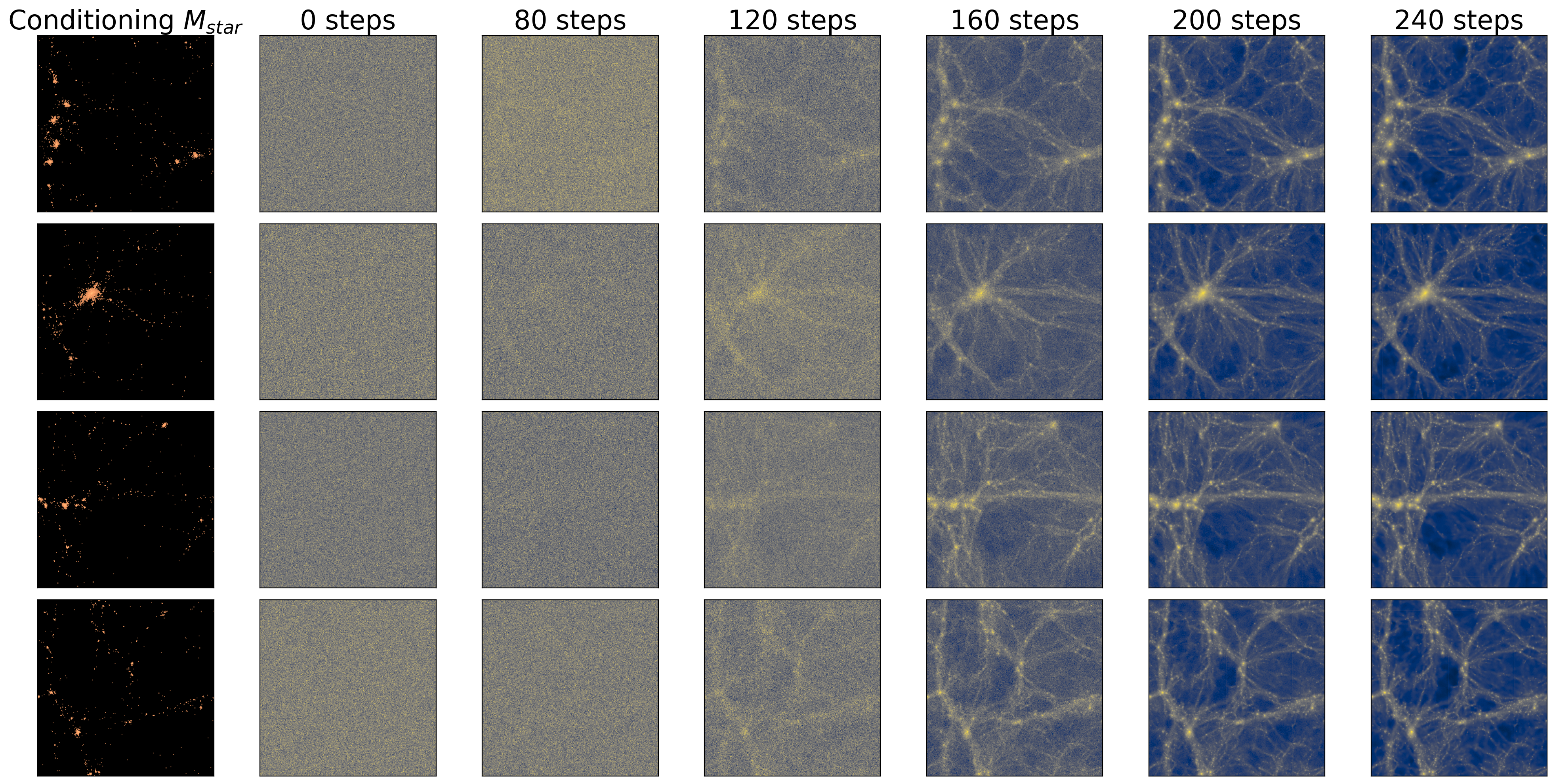}
    \caption{Input conditioning and sampled images at different steps of the diffusion process. Visually, most of the structure are produced between steps 80 and 160.}
    \label{fig:supp_diffusion}
\end{figure}

We show the diffusion process generating DM fields conditioned on stellar mass fields in Figure \ref{fig:supp_diffusion}. Visually, we find that most of the structure forms between steps 80 and 160 when discretizing $t=[0,1]$ into 250 steps and a learned noise schedule of $\gamma(t)=18.347\:t-7.202$. We hope to investigate if there could be a better noise scheduling for our task which can improve the results.

\begin{figure}
    \centering
    \includegraphics[width=\textwidth]{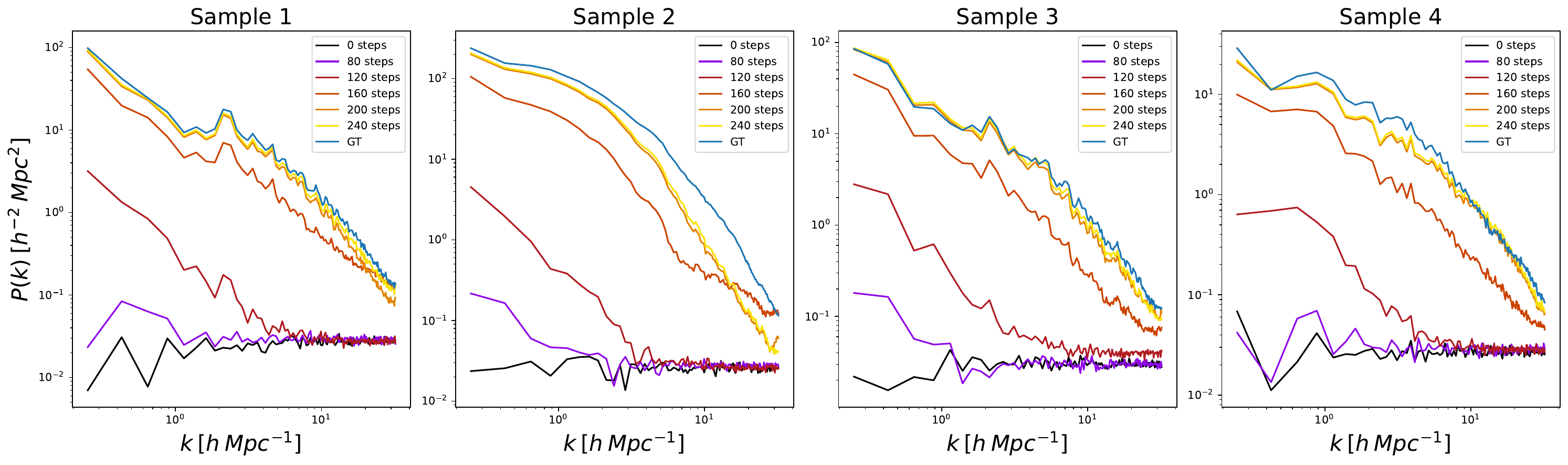}
    \caption{The evolution of the powerspectrum during the diffusion process. We show the powerspectrum of the DM field in different steps of the diffusion process for 4 randomly chosen stellar mass inputs.}
    \label{fig:supp_pk_evol}
\end{figure}

In Figure \ref{fig:supp_pk_evol}, we show for 4 input stellar mass fields the evolution of the power spectrum in the diffusion process. The diffusion process seems to produce the large scale features first before completing the small scales.

\subsection{Compute Time}
We used a single NVIDIA A100 40 GB GPU for training and evaluation. With a batch size of 12, each training step took $\sim$ 0.2 s and sampling a batch of images with 250 steps took 8.4 s.


\end{document}